# Stock market forecasting using DRAGAN and feature matching

Fateme Shahabi Nejad [a], Mohammad Mehdi Ebadzadeh [a,*]

[a] *Department of Computer Engineering and Information Technology, Amirkabir University of Technology (Tehran Polytechnic), Tehran, Iran.*

## ABSTRACT

Applying machine learning methods to forecast stock prices has been one of the research topics of interest in recent years. Almost few studies have been reported based on generative adversarial networks (GANs) in this area, but their results are promising. GANs are powerful generative models successfully applied in different areas but suffer from inherent challenges such as training instability and mode collapse. Also, a primary concern is capturing correlations in stock prices. Therefore, our challenges fall into two main categories: capturing correlations and inherent problems of GANs. In this paper, we have introduced a novel framework based on DRAGAN[1] and feature matching for stock price forecasting, which improves training stability and alleviates mode collapse. We have employed windowing to acquire temporal correlations by the generator. Also, we have exploited conditioning on discriminator inputs to capture temporal correlations and correlations between prices and features. Experimental results on data from several stocks indicate that our proposed method outperformed long short-term memory (LSTM) as a baseline method, also basic GANs and WGAN-GP[2] as two different variants of GANs.

*Keywords:* stock price prediction, time series forecasting, Generative Adversarial Networks, feature matching.

## 1. Introduction

Stock price forecasting is a challenging task because it is affected by various factors such as economic components, political issues, the financial performance of the company, etc. Traditional methods of stock price forecasting can be classified into two general categories: technical analysis and fundamental analysis. The technical analysis predicts the future stock price using historical data such as stock price and trading volume and captures the future trends of the stock market. Among the prevalent methods of technical analysis, we can mention moving average (MA), technical indicators, support and resistance levels, trend lines, candlestick patterns, and channels. Fundamental analysis typically deals with qualitative data like financial news and economic data. It evaluates the company's stock based on its intrinsic value, financial reports, market situation,

* Corresponding author
*Email addresses:* fa.shahabi@aut.ac.ir (Fateme Shahabi Nejad) , ebadzadeh@aut.ac.ir (Mohammad Mehdi Ebadzadeh )

*Preprint submitted to Elsevier*

[1] Deep Regret Analytic Generative Adversarial Network
[2] Wasserstein GAN – Gradient Penalty

etc. Fundamental analysis is a proper methodology for long-term investment and commonly does not change with short-term news. In contrast, technical analysis is more suitable for short-term trading, and news usually affects it.

Applying machine learning methods to forecast stock prices is one of the research topics of interest in recent years. Among them, GANs have a high potential in learning the distribution of high-dimensional and complex data. These networks do not consider any specific assumption about the distribution and can generate real-like samples from a hidden space based on the adversarial approach. This powerful characteristic leads to their use in various applications. In this paper, we have proposed a novel framework based on DRAGAN, which forecasts stock prices based on historical data and technical indicators. The most important contributions of this paper are as follows:

- To our knowledge, this paper is the first to employ DRAGAN and feature matching for forecasting stock prices.
- We use feature matching and customized cost functions to design the proposed framework, which mitigates mode collapse[3] and improves training stability.
- We found that employing windowing and conditioning in our method can better capture temporal correlations and correlations between prices and features.
- Our method is compared with LSTM as a baseline method, also basic GANs and WGAN-GP as two different variants of GANs. Experiments show that it performs better than the compared methods. Indeed, the proposed approach can better learn the future trends of stock.
- Since the design of the proposed framework intends to capture correlations, we may be able to successfully apply it to similar tasks in the time series area.

The organization of this paper is as follows: Section 2 deals with an overview of the research conducted for stock price forecasting based on the machine learning attitude. Section 3 provides a brief discussion of the theoretical concepts of the paper. Section 4 first introduces the overall structure of the proposed framework and then discusses its components in more detail. The experimental results of our proposed method and its comparison with baseline methods are presented in section 5. Finally, concluding remarks and suggestions for future research are given in Section 6.

**2. Related works**

Most recently, the popularity of employing machine learning methods to forecast stock prices has been increasing. In this regard, various techniques have been used for stock forecasting, ranging from regression methods to artificial neural networks (ANN). Most of these methods rely on the combination of machine learning and technical analysis, and fundamental analysis has rarely been used. The reason is that technical data is available daily and even hourly, while fundamental data is reported quarterly and annually. Also, the nature of technical data is quantitative and structured. On the other side, by considering the stock price as time series, some classic time series algorithms have also been extended to forecast the stock price. By reviewing the conducted studies for stock price forecasting, we have summarized them into the following three categories: classic methods, traditional machine learning methods, and deep learning methods.

---

[3] The mode collapse problem occurs when the generator maps different inputs to the same output. In other words, the generator produces only a particular sample or a limited variety of samples.

**Classic methods:** By considering the stock price as time series, some statistical time-series methodologies such as auto-regression (AR), auto-regressive moving average (ARMA), and auto-regressive integrated moving average (ARIMA) have been developed to forecast the stock price (Box et al., 2015). Due to the non-linearity of stock data, ARIMA does not perform well on its own. Areekul et al. (2009) employed a neural network to improve the performance of ARIMA. The combination of ARIMA and support vector machine (SVM) was suggested for stock price forecasting by Pai and Lin (2005). Commonly, classic methods have poor results compared to the other two categories.

**Traditional machine learning methods:** Traditional approaches include methods such as naive Bayes, SVM, random forest, k-nearest neighbor (KNN), fuzzy theory, and linear regression. Fenghua et al. (2014) decomposed the stock price into features using singular spectrum analysis (SSA). Then an SVM applied these features for price prediction. Patel et al. (2015) evaluated the performance of four machine learning methods, including ANN, SVM, random forest, and naive Bayes, in forecasting the Indian stock market. Bhuriya et al. (2017) applied linear regression to predict the stock market and found that it outperformed the polynomial and radial basis function (RBF) regression models. Ebadati and Mortazavi (2018) suggested a combined method based on genetic algorithms and ANN to make stock price predictions. Sinaga et al. (2019) proposed an effective mixture model for stock trend prediction. They modified support vector k-nearest neighbor clustering (SV-kNNC) by replacing the k-means clustering with a self-organizing map (SOM). Zhang et al. (2019) presented a new prediction method using fuzzy time series with a genetic algorithm. Sedighi et al. (2019) introduced a novel prediction model that uses an adaptive neuro-fuzzy inference system (ANFIS), an artificial bee colony algorithm, and an SVM. Wang et al. (2019) extended a new hybrid model based on fractal interpolation and SVM models to forecast time series of stock price indexes. Basak et al. (2019) facilitated random forest and gradient boosted decision trees (using XGBoost) to predict stock price direction. Cao et al. (2019) utilized a complex network method. Then network characteristic variables are applied as input variables for KNN and SVM methods to predict the stock price patterns. Nasiri and Ebadzadeh (2022) proposed a multi-functional recurrent fuzzy neural network (MFRFNN) based on two neural networks for chaotic time series prediction. They compared the proposed model with the decision tree, random forest, support vector regression (SVR), multi-layer perceptron (MLP), and LSTM.

**Deep learning methods:** With the development of deep learning, stock price forecasting using it has received much attention. Deep learning can learn complex nonlinear structures and suitable features without the need for the prior knowledge of experts (Chong et al., 2017). Singh and Srivastava (2017) employed principal component analysis (PCA) with a deep neural network (DNN) and demonstrated that deep learning could improve the accuracy of stock price forecasting.

Convolutional neural network (CNN) is one of the most successful types of networks in deep learning, which has extensive applications in image processing, natural language processing, speech recognition, etc. Cao and Wang (2019) made stock index forecasting using a CNN and a CNN+SVM. Empirical results indicated that two prediction models were efficient.

LSTM network has promising results for time series forecasting. In this context, many studies have been reported based on LSTM or hybrid with it. Di Persio and Honchar (2017) compared the performance of different recurrent neural networks (RNNs), including the basic multi-layer RNN, LSTM, and gated recurrent unit (GRU). Their results for predicting Google stock price showed that LSTM performed better than other variants. Fischer and Krauss (2018) applied LSTM networks to a large-scale financial market prediction task on the S&P 500. They demonstrated that it performed better than random forest, DNN, and logistic regression. Chen and Ge (2019) found that using the attention mechanism could improve the performance of the LSTM-based prediction method. Jin et al. (2020) suggested investors' sentiment for stock prediction, which led to improved performance. Also, they applied an attention mechanism in combination with LSTM to extract important information. Xu et al. (2020) employed the wavelet transform to reduce the noise of stock data and a stacked auto-encoder to remove unessential features. Finally, a bidirectional LSTM was applied to predict the stock price. Gunduz (2021) utilized LSTM with a variational auto-encoder (VAE) to predict eight banking stocks in Borsa Istanbul.

The accurate performance of deep networks usually depends on the amount of training data. So, several studies have utilized transfer learning to overcome insufficient training samples and overfitting problems. Nguyen and Yoon (2019) leveraged the advantages of transfer learning in combination with LSTM to mitigate the above challenges. Gu and Dai (2021) presented a new multi-source transfer learning algorithm to address the problems of insufficient training data and differences in the old and new data distribution.

Several studies have attempted to adapt GANs for the task of stock price prediction. Zhou et al. (2018) proposed basic GANs with LSTM as generator and CNN as a discriminator for adversarial forecasting of stock prices. They modified the loss function of basic GANs by adding direction prediction loss and forecast error loss. Their experiments showed that GANs outperformed other compared models such as ARIMA, ANN, SVM, and LSTM. Zhang et al. (2019) suggested basic GANs with LSTM as the generator and MLP as the discriminator to predict closing price. They demonstrated that GANs had better results than SVR, ANN, and LSTM. Sonkiya et al. (2021) made a sentiment analysis of the news for Apple Inc. Then, basic GANs predicted the stock price based on sentiment scores and some features. The proposed model performed better than benchmark models like ARIMA, GRU, LSTM, and basic GANs (without sentiment analysis). Lin et al. (2021) used news sentiment analysis and WGAN-GP to predict stock prices. They employed GRU for the generator and CNN for the discriminator. Examining the results on Apple stock showed that WGAN-GP had better performance than LSTM, GRU, and basic GANs.

A review of studies demonstrates that deep learning methods have been more successful than traditional and classic methods (Bustos & Pomares-Quimbaya, 2020; Kumbure et al., 2022). Among them, LSTM is the most widely used method for stock price forecasting. Due to the difficulties of GANs in adapting to the task of stock price forecasting, relatively few studies have been reported based on them, but their results are promising. So in this paper, we have proposed a novel GANs-based method and compared its results with successful models such as LSTM, WGAN-GP, and Basic GANs.

## 3. Backgrounds

### 3.1 GANs

After the introduction of GANs in 2014 (Goodfellow et al., 2014), the employment of these models in various fields of machine learning has rapidly extended. GANs consist of two models, the generator and the discriminator. The generator G aims to produce real-like samples that the discriminator cannot recognize them as fake. The discriminator D also intends to distinguish real samples from fake ones. In an adversarial learning process, the two models reach an equilibrium point where the generator can capture the distribution of real data, and the discriminator is unable to distinguish fake and real samples from each other. The training process includes simultaneous optimization of D and G in a two-player game. The value function for the basic GAN is defined as follows:

$$\min_G \max_D V(D, G) = E_{x \sim P_r}[\log D(x)] + E_{z \sim p_z(z)}[\log(1 - D(G(z)))] \quad (1)$$

where $x$ are samples of real data under the $P_r$ distribution. The generator takes the vector of random noise $z \sim p_z(z)$ as input and produces fake samples $G(z)$. $D(.)$ is also the output of the discriminator, which expresses the probability of the input being real as a single scalar.

### 3.2 Long Short-Term Memory (LSTM)

LSTM models are a particular type of RNN that are able to learn long-term dependencies and have been successful in time series processing. Indeed, Hochreiter and Schmidhuber (1997) proposed LSTM to mitigate the problems of vanishing and exploding gradients associated with RNN. The structure of an LSTM cell consists of three components: the forget gate, the input gate, and the output gate. The forget gate determines whether information should be preserved or discarded. The input gate controls what new information can be added to the cell. Finally, the output gate decides what data to send to the next cell.

### 3.3 Gated Recurrent Units (GRUs)

GRUs are a simplified version of LSTM introduced by Cho et al. (2014). Because GRUs only include the reset gate and the update gate, they are faster than LSTMs. The reset gate determines how much of the previous information to forget and can capture short-term dependencies. The update gate has a function similar to the forget gate and the input gate of an LSTM. It decides what information to discard and what new information to add. The update gate can capture long-term dependencies. We cannot express an undoubted opinion on the preference of LSTM or GRUs. Researchers usually examine both models and choose the appropriate model depending on their application.

## 4. Our Methodology

In this section, the details of the proposed framework for stock price forecasting are examined. We intend to forecast the closed price of a particular stock based merely on its historical data and some technical indicators. Our primary investigations indicated that applying naive GANs to time series gives undesirable results. In other words, naive GANs are unable to capture temporal correlations and correlations between price and features. Moreover, GANs suffer from inherent challenges such as training instability and mode collapse. Therefore, our challenges fall into two main categories: capturing correlations and inherent problems of GANs. In this paper, a novel framework has been proposed to mitigate these challenges. The overall structure of this framework can be seen in Fig. 1.

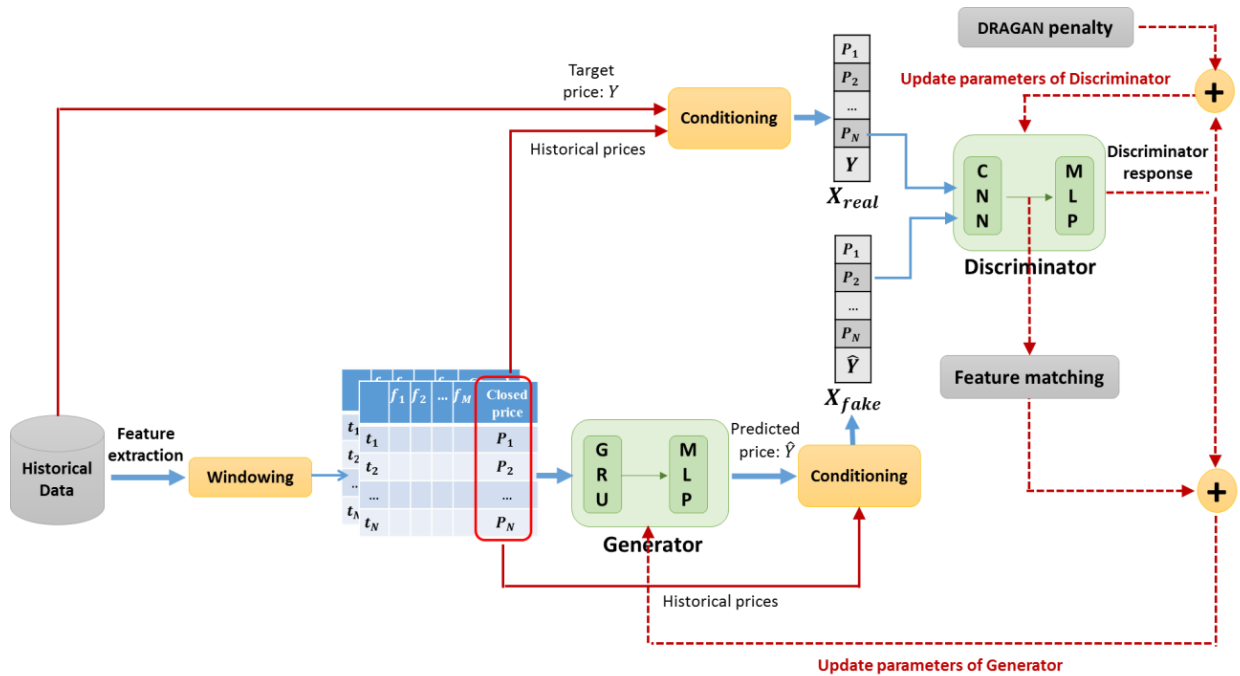

**Fig. 1.** The overall structure of our proposed framework.

First, the desired features are extracted from the historical data. The combination of historical data and technical indicators makes the features denoted by $f_1, f_2, ..., $ and $f_M$ in Fig. 1. Afterward, the acquired data is segmented based on the windowing process and fed to the generator as input. The generator outputs the predicted price as $\hat{Y}$. In the conditioning step, historical prices and predicted price are concatenated, and the result is fed to the discriminator as input $X_{fake}$. Historical prices are the closed prices for N consecutive time steps in a segment, where N is the size of the segment (or the window length). Similarly, the concatenation of historical prices and the target price is fed to the discriminator as input $X_{real}$. The generator and discriminator are adversarially trained based on our customized cost functions. In the following, we will examine the details of our framework.

## 4.1 Windowing

The windowing technique is applied to capture temporal correlations and correlations between prices and features. That is, a window is rolled over the training set and test set, and segments of the time series with a certain length are acquired. Each segment contains historical price information and features for several consecutive time steps. So, we feed these segments rather than a single time step as an input to the generator. Indeed, the generator can learn many-to-one or many-to-many mapping instead of one-to-one mapping and predict the future price based on segments of historical prices and features. For example, the model takes a three-day time series as input and predicts the next two days.

## 4.2 Conditioning

The second approach to capture correlations is to use conditional GANs. Mirza and Osindero (2014) introduced a conditional version of GANs in which both the generator and discriminator are conditioned on some secondary information. This secondary information can be any information such as class labels or other auxiliary data. We can perform the conditioning by feeding secondary information into the discriminator and generator. In this paper, conditioning means that historical prices are concatenated with the predicted price and sent to the discriminator as input. Another discriminator input is fed from the concatenation of historical prices and target price. Here, historical prices are equivalent to secondary information. Using conditioning can help to capture temporal correlations and correlations between prices and features and improve forecasting accuracy.

## 4.3 Applying DRAGAN

GANs are powerful generative models but suffer from training instability and mode collapse. Many advances have been made to address these challenges during these years. We investigated how to leverage known state-of-the-art GANs to facilitate stock forecasting. WGAN[4] is one of these popular models introduced by Arjovsky et al. (2017). They replaced the Jensen-Shannon divergence used in the basic GANs with the Earth-Mover (also called Wasserstein) distance. The Earth-Mover distance is the minimum cost of transporting mass in order to transform the source distribution into the destination distribution (Arjovsky et al., 2017). The cost is the mass multiplied by the transport distance. The WGAN cost function is simplified using the Kantorovich-Rubinstein duality (Villani, 2009) as follows:

$$\min_{G} \max_{D \in \mathcal{D}} \mathop{\mathrm{E}}_{X_{real} \sim P_r}[D(X_{real})] - \mathop{\mathrm{E}}_{X_{fake} \sim P_g}[D(X_{fake})] \qquad (2)$$

where $P_g$ and $P_r$ represent the generated and real data distributions, respectively. Here, the discriminator is called a critic because it outputs a scalar score rather than a probability. This score indicates how real the input is. The critic response is denoted by $D(X)$, and $\mathcal{D}$ is the set of 1-Lipschitz functions. A differentiable function is known as 1-Lipschitz if and only if it has gradients with the norm at most one everywhere (Gulrajani et al., 2017). To make the 1-Lipschitz constraint on the critic, WGAN clips the weights to fall within the specified range [−c, c]. Weight clipping sometimes leads to optimization problems and non-convergence. So, WGAN-GP proposed another way to enforce the 1-Lipschitz constraint (Gulrajani et al., 2017). It penalizes the critic if the gradient norm deviates from the value of 1. The WGAN-GP cost function is defined as follows:

$$\min_D \mathop{E}_{X_{fake} \sim P_g}[D(X_{fake})] - \mathop{E}_{X_{real} \sim P_r}[D(X_{real})] + \lambda \cdot \mathop{E}_{\hat{X} \sim P_{\hat{X}}}\left[(\|\nabla_{\hat{X}} D(\hat{X})\|_2 - 1)^2\right] \quad (3)$$

$$\hat{X} \leftarrow \epsilon X_{real} + (1 - \epsilon)X_{fake} \,, \; \epsilon \sim U[0, 1]$$

where $\hat{X}$ is any point interpolated based on samples from $P_g$ and $P_r$. Experiments show that WGAN-GP improves training stability and performs better than basic GANs and WGAN.

---

[4] Wasserstein GAN

DRAGAN is one of the well-known cost functions for training GANs (Kodali et al., 2017). The authors suggested a new perspective to investigate GAN training based on regret minimization. According to their theory, mode collapse is due to trapping at undesirable local equilibria. In situations of mode collapse, gradients of the discriminator around some real data points are sharp. Hence, a novel gradient penalty is proposed that penalizes the discriminator gradient near the real data points:

$$\lambda \cdot E_{X_{real} \sim P_r, \delta \sim N_d(0,cI)} \; [\|\nabla_X D_\theta(X_{real} + \delta)\|_2 - k]^2 \quad (4)$$

where $X_{real}$ is the real point, and $X_{real} + \delta$ denotes the noisy sample. Their evaluation showed that DRAGAN mitigated the mode collapse issue and improved stability.

The difference between WGAN-GP and DRAGAN lies in where the gradient penalty term is applied. When the generator does not perform well (the generated samples are far from the real samples), the gradient penalty calculation based on WGAN-GP has more disadvantages due to poor interpolation. We have taken advantage of the DRAGAN idea in designing our framework and added its penalty term to our cost function. Our explorations show that it leads to more stable training and performs better than WGAN-GP.

**4.4 Feature Matching**

GANs training intends to find the Nash equilibrium[5] of a non-convex game based on two agents. Gradient descent methods are proper for solving traditional optimization problems based on a single agent rather than multiple agents. Therefore, employing these methods for training GANs may lead to non-convergence (Arjovsky & Bottou, 2017). Unfortunately, there is no suitable alternative method, and some researchers only have suggested techniques to alleviate these challenges. Salimans et al. (2016) proposed several techniques to improve the convergence of the GANs game. These techniques are based on a heuristic understanding of the non-convergence problem. Among them, we have utilized feature matching to make our generator cost function more stable. For this purpose, the following expression is added to the generator cost function:

---

[5] According to game theory, the convergence of GANs occurs at the Nash equilibrium point. The Nash equilibrium is the optimal point for the min-max equation of GANs. That is, no player can decrease their cost function without changing the parameters of other players.

$$\left\| E_{X_{real} \sim P_r} f(X_{real}) - E_{X_{fake} \sim P_g} f(X_{fake}) \right\|_2^2 \qquad (5)$$

Here, $f(x)$ indicates the output of a specified intermediate layer of the discriminator for the input $x$. Since the discriminator is trained to distinguish real data from generated data, $f(x)$ is a proper representation of the features.

Feature matching intends to minimize the difference between features of real and generated data. Our examinations show that feature matching can mitigate the unstable behavior of GANs and lead to more forecasting accuracy. Indeed, it conducts the generator to produce data that matches the statistics of the real data. Instead of relying only on the last layer output of the discriminator, the generator also considers the intermediate layers and thus can avoid overtraining. As shown in Fig. 1, we simultaneously leverage feature matching and discriminator response to design our generator's cost function.

### 4.5 Cost Functions and Networks Architecture

Proper design of cost functions, generator architecture, and discriminator architecture are fundamental components in the success of the GANs-based model. The generator aims to learn the stock price distribution. So, it takes a window of consecutive data and outputs the predicted price. As mentioned earlier, LSTM and GRU are able to model time series correlations well. Our preliminary evaluations show that GRU performs better than LSTM in our framework and is also faster. Therefore, we use two layers of GRU for the generator architecture with the number of 256 and 128 units, respectively. Next, two fully connected layers are added on top of this architecture to map the data of GRU layers to the predicted prices. The purpose of the discriminator is to classify real data from generated data. It takes real and generated data as inputs and returns a scalar score that determines how real the input is. Due to the success of CNN in classification, our discriminator architecture consists of three convolutional layers with 32, 64, and 128 units, respectively. Also, two fully connected layers are placed on top of them.

We already explained that DRGAN improves training stability and alleviates mode collapse issues. Therefore, we utilize the combination of the DRAGAN penalty term and the Wasserstein distance to define our discriminator cost function. These two terms are added together with the weighting parameter $\lambda_1$ as follows:

$$\min_D \ \mathop{E}_{X_{fake} \sim P_g}[D_\theta(X_{fake})] - \mathop{E}_{X_{real} \sim P_r}[D_\theta(X_{real})] + \lambda_1 \cdot E_{X_{real} \sim P_r, \delta \sim N_d(0,cI)} \left[ \| \nabla_X D_\theta(X_{real} + \delta) \|_2 - k \right]^2 \qquad (6)$$

We construct our generator cost function by adding the feature matching expression to the discriminator response:

$$\min_G \ -\mathop{E}_{X_{fake} \sim P_g}[D_\theta(X_{fake})] + \lambda_2 \cdot \left\| E_{X_{real} \sim P_r} f(X_{real}) - E_{X_{fake} \sim P_g} f(X_{fake}) \right\|_2^2 \qquad (7)$$

Here, the $\lambda_2$ parameter determines the weight of feature matching. In Fig. 1, dashed lines indicate the backpropagation of gradients corresponding to our cost functions.

## 5. Experiments

We evaluate our proposed framework on data from several stocks, including Apple, Microsoft, Amazon, Nvidia, Google, and Tesla. This data is collected from Yahoo Finance and contains historical stock prices from 2010 to 2020. After feature extraction, the set of 14 features given as input to the generator is as follows:

- Historical data on the stock include High, Low, Open, Close, Adjacent close, and Volume.
- Some popular technical indicators are calculated based on stock price data. Among them, we can mention MA7 (7-day moving average), MA21 (21-day moving average), MACD (moving average convergence/divergence), EMA (exponential moving average), logarithmic momentum, and Bollinger Bands.

All the above features are normalized to map their values to the range of 1 and -1. We consider 70% of the data for training and 30% for testing. Then, the windowing step is performed separately for training and testing data. For this purpose, a window with a length of 3 is rolled over the data, and segments with a length of 3 time steps are obtained. Each time step includes all the mentioned features. So, segments with dimensions of 3*14 are fed to the generator as input, where 14 is the number of features.

Our goal is to forecast the next day's price based on the information from three consecutive days. Hence, the generator outputs the price for the next day. In the conditioning step, we concatenate the predicted price with the price of three consecutive days and feed it to the discriminator as fake input. Also, we provide the concatenation of the target price with the price of three consecutive days to the discriminator as real input. The target price is the true price of the next day and can be extracted from the training and test sets.

During adversarial training, both the generator and the discriminator optimize their cost function alternatively to converge to the equilibrium point. The model is trained using Adam optimizer based on the proposed cost functions in Eqs. (6) and (7). We use $\lambda_1 \sim 10$, $\lambda_2 \sim 1$, $k = 1$, and $c \sim 10$ as optimal values of the hyperparameters in our experiments. Once trained, we solely employ the generator as the final forecasting model. We evaluate the performance of each model by root mean square error (RMSE), which is a proper metric for regression problems and is defined as:

$$RMSE = \sqrt{\frac{\sum_{i=1}^{N}(Y_i - \hat{Y}_i)^2}{N}} \quad (8)$$

Here, N denotes the number of data points. $\hat{Y}$ and $Y$ refer to the predicted price and target price, respectively. The proposed model is compared with LSTM as a baseline, basic GANs, and WGAN-GP. The LSTM model includes a bidirectional LSTM layer with 128 units and a fully connected layer on top of it. To ensure a fair comparison, we employ windowing, conditioning, and the same architecture across all GANs-based models (as described in Section 4.5). The cost functions for training the basic GANs and WGAN-GP are based on Eqs. (1) and (3), respectively.

The evaluation results for all methods are summarized in Table 1. According to these results, our proposed model outperforms baseline approaches on all considered stocks. WGAN-GP commonly arrives in second place in terms of performance, but sometimes basic GANs have better results with a slight difference (Amazon and Tesla stocks). It is noteworthy that for a stock like Google that has more fluctuations, LSTM performs better than basic GANs.

The results of the predicted prices on the test set are depicted as stock charts in Fig. 2, which are consistent with the RMSE values. These charts show that the proposed method can acquire the stock trends correctly.

**Table 1.** The RMSE Results of our proposed model compared to the baseline methods (prediction 3 to 1).

| Stocks | | Our Model | WGAN-GP | Basic GANs | LSTM |
|---|---|---|---|---|---|
| Apple | Train | **0.458** | 0.508 | 0.587 | 0.612 |
| | Test | **1.047** | 1.257 | 1.691 | 1.741 |
| Microsoft | Train | 0.871 | 1.462 | **0.856** | 0.992 |
| | Test | **2.143** | 2.835 | 3.306 | 3.738 |
| Amazon | Train | 0.637 | 0.729 | **0.617** | 0.625 |
| | Test | **2.068** | 2.161 | 2.121 | 2.232 |
| Nvidia | Train | **0.403** | 1.314 | 0.703 | 0.521 |
| | Test | **2.124** | 2.347 | 2.591 | 2.698 |
| Google | Train | 0.594 | 1.818 | 0.604 | **0.518** |
| | Test | **1.489** | 1.491 | 1.667 | 1.537 |
| Tesla | Train | 0.439 | 0.536 | 0.606 | **0.427** |
| | Test | **0.826** | 0.968 | 0.955 | 1.233 |

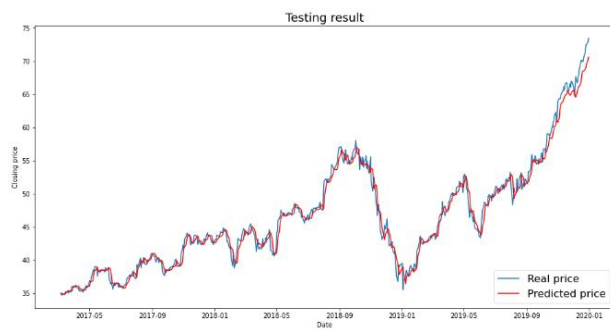

(a)

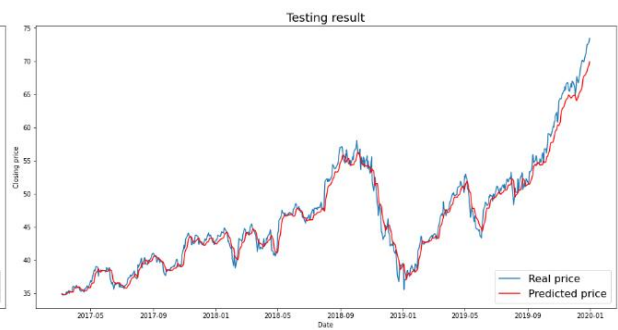

(b)

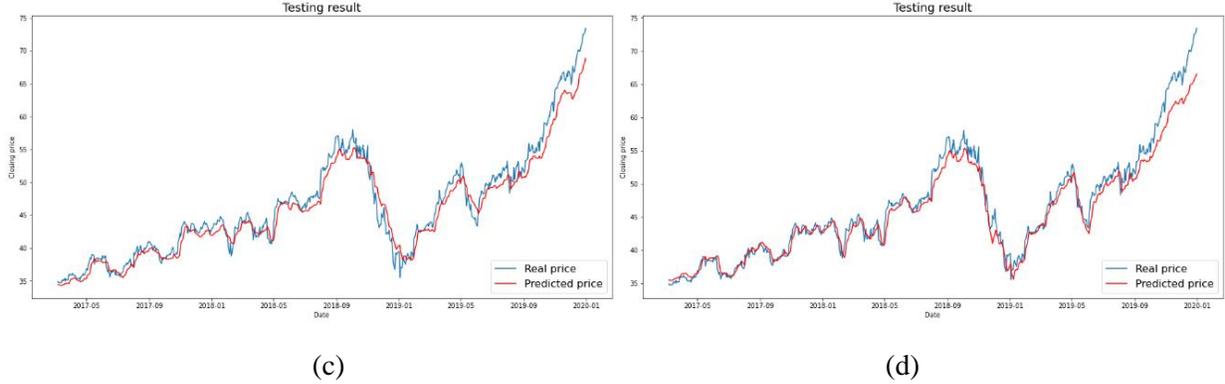

|     (c)     |     (d)     |

**Fig. 2.** Comparison of the predicted price and real price based on the test set (Apple stock). (a) Our proposed model. (b) WGAN-GP. (c) Basic GANs. (d) LSTM.

We have also conducted another experiment to investigate the ability of the proposed model for many-to-many mapping. For this purpose, we have considered the size of the input window to the generator as 10, and the goal is to forecast the stock price in the coming days based on the data of 10 consecutive days. This experiment has been executed to predict the price of 1, 2, 3, and 5 days ahead. As shown in Table 2, our model is able to learn many-to-many mapping better than the compared methods. These results show that LSTM is ranked second in terms of performance (except for the 10 to 5 mapping). In addition, comparing the results of Tables 1 and 2 indicates that LSTM can learn long-term dependencies better than the basic GANs and WGAN-GP methods.

**Table 2.** Comparison of the results for many-to-many mapping on Apple stock.

| Method | Training RMSE | | | | Testing RMSE | | | |
| --- | --- | --- | --- | --- | --- | --- | --- | --- |
|  | 10 to 1 | 10 to 2 | 10 to 3 | 10 to 5 | 10 to 1 | 10 to 2 | 10 to 3 | 10 to 5 |
| Our Model | 0.756 | 0.985 | **0.732** | 0.808 | **1.133** | **1.561** | **1.625** | **1.715** |
| WGAN-GP | 0.778 | 0.832 | 0.882 | 0.863 | 1.438 | 1.727 | 1.904 | 2.081 |
| Basic GANs | 0.804 | 0.763 | 0.864 | 0.851 | 1.821 | 1.934 | 1.996 | 2.271 |
| LSTM | **0.713** | **0.633** | 0.776 | **0.790** | 1.311 | 1.716 | 1.880 | 2.234 |

We intended to compare our model with the results of previous studies, but the direct comparison was not reasonable due to the unavailability of their code details, using different datasets, and different time frames. We have summarized the results of popular GANs-based studies by methodology, type of features, dataset, and reported experimental results in Table 3. To the best of our knowledge, WGAN-GP has performed better than other GANs variants in these studies, while we found that our model indicates better prediction results than WGAN-GP. Sonkiya and Lin's studies employed sentiment analysis in combination with GANs models to forecast Apple's stock price over the same time period. Comparing their results with our method on the same stock indicates that our proposed method is promising. So, it can be improved by utilizing sentiment analysis and more efficient features.

From another point of view, we have investigated the ability of our model to learn real data distribution. Our explorations show that the distributions of predicted and real prices are very close to each other, and our model has performed well in learning real data distribution. An example of this comparison is presented in Fig. 3. Finally, we can see the price prediction charts for every six stocks using the proposed model in Fig. 4.

**Table 3.** Summary of common GANs-based studies for stock price forecasting.

| Model | Methodology | Features | Dataset | Results |
|---|---|---|---|---|
| Zhou et al. (2018) | Basic GANs with LSTM as generator and CNN as discriminator. | Historical prices + technical indicators. (13 features) | 42 stocks in China stock market. Jan. 2016 - Dec. 2016. Minute time frame. | Different values based on metrics other than RMSE. |
| Zhang et al. (2019) | Basic GANs with LSTM as generator and MLP as discriminator. | Historical prices + MA5. (7 features) | S&P 500, IBM, Microsoft, Shanghai Composite Index, Ping An Insurance Company of China (PAICC). 1998 – 2018. Daily time frame. | RMSE: 4.103 on Avg. |
| Sonkiya et al. (2021) | Basic GANs with GRU as generator and CNN as discriminator + sentiment analysis. | Historical prices + technical indicators + stock indexes of some countries + several commodities. (37 features) | Apple. July 2010 - July 2020. Daily time frame. | RMSE: 1.827 |
| Lin et al. (2021) | WGAN-GP with GRU as generator and CNN as discriminator + sentiment analysis. | Historical prices + technical indicators + stock indexes of some countries + several commodities. (36 features) | Apple. July 2010 - July 2020. Daily time frame. | RMSE: 3.880 |
| Our proposed model | DRAGAN with GRU as generator and CNN as discriminator + feature matching. | Historical prices + technical indicators. (14 features) | Apple, Microsoft, Amazon, Nvidia, Google, and Tesla. July 2010 - July 2020. Daily time frame. | RMSE: 1.044 on Apple 1.616 on Avg. |

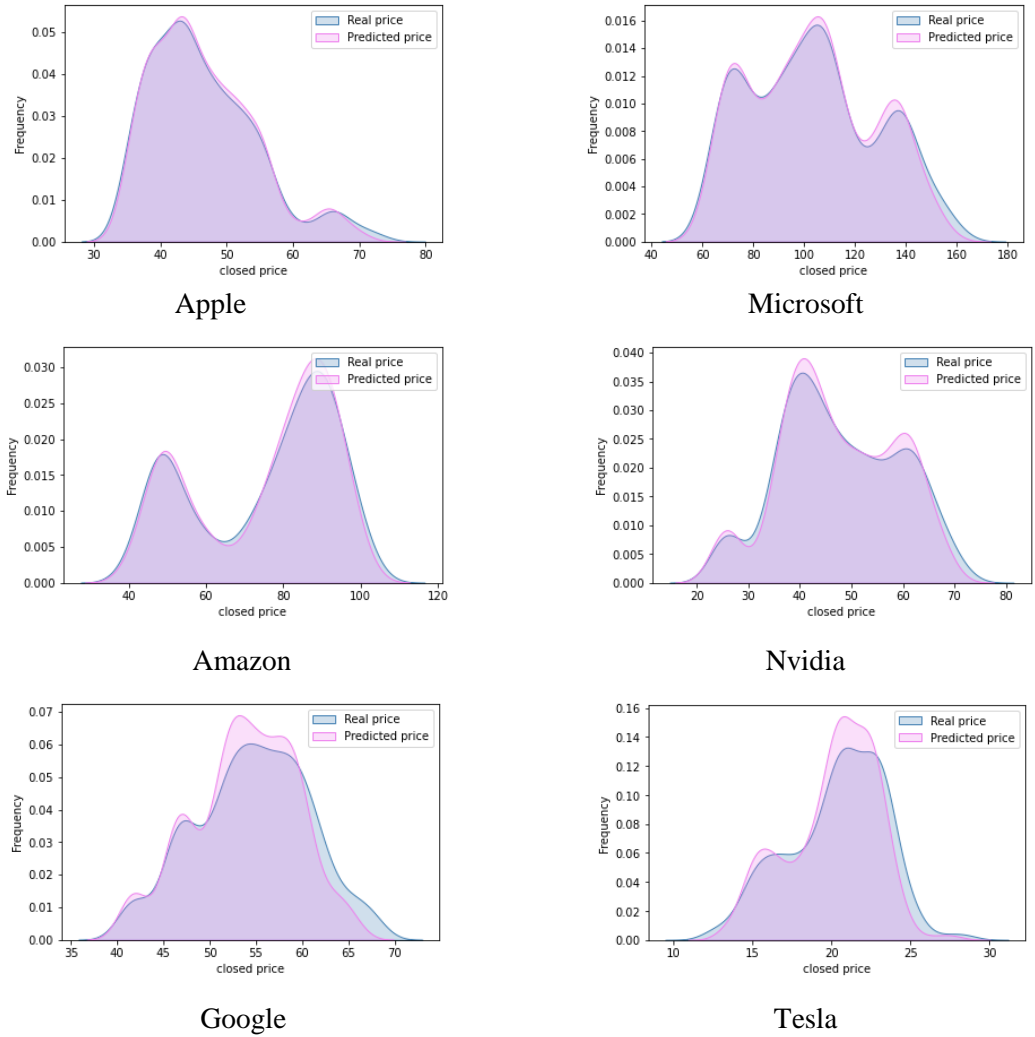

**Fig. 3.** Predicted prices distribution versus real prices distribution.

## 6. Conclusion

In this paper, we have presented a novel framework based on DRAGAN and feature matching for stock price forecasting that improves training stability and mitigates mode collapse. We have employed windowing to acquire temporal correlations by the generator. The second approach to capture correlations is to use conditional GANs. We have exploited conditioning on discriminator inputs to capture temporal correlations and correlations between prices and features. In addition, we have proposed customized cost functions for training our generator and discriminator. Our experiments on data from several stocks demonstrate that the proposed method performs better than other baseline methods such as LSTM, basic GAN, and WGAN-GP.

Based on our knowledge, the few GAN-based studies in this area are mainly basic GANs and WGAN variants. Indeed, training instability and non-convergence are the main barriers to developing GANs for stock price forecasting. Therefore, the promising results of our approach can be a motivation for future research. In addition, we need to explore GANs-based methods on more extensive and diverse data sets. It is worth mentioning that a limited number of studies have employed sentiment analysis of the news in combination with GANs models, but their results need further examination due to their insufficient data (Lin et al., 2021; Sonkiya et al., 2021). Therefore, one suggestion to improve our model is to investigate the use of sentiment analysis and also focus on producing more appropriate features.

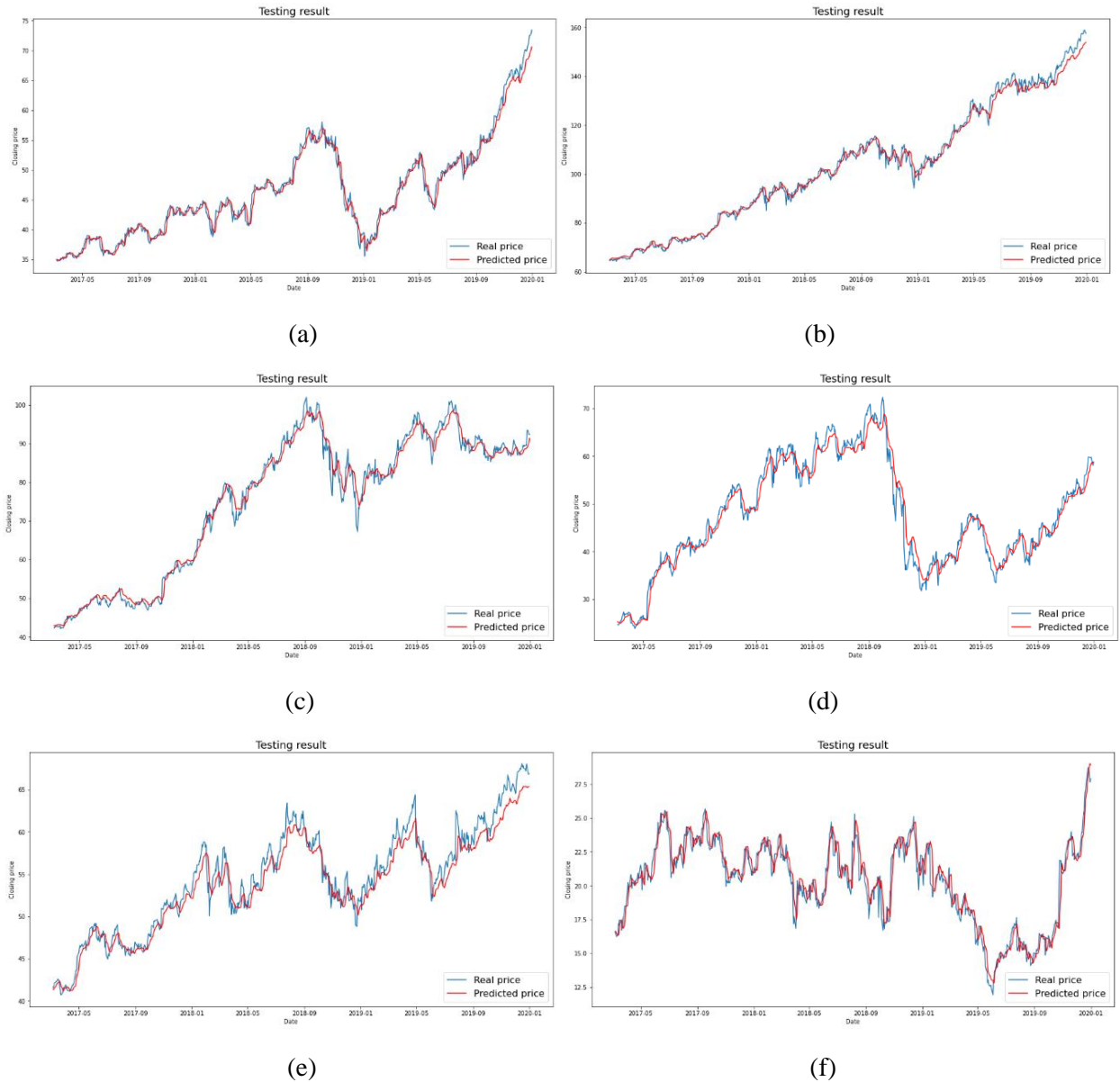

**Fig. 4.** The price prediction charts using our proposed model for various stocks. (a) Apple. (b) Microsoft. (c) Amazon. (d) Nvidia. (e) Google. (f) Tesla.

# References

Areekul, P., Senjyu, T., Toyama, H., & Yona, A. (2009). A hybrid ARIMA and neural network model for short-term price forecasting in deregulated market. *IEEE Transactions on Power Systems*, *25*(1), 524–530.

Arjovsky, M., & Bottou, L. (2017). Towards principled methods for training generative adversarial networks. *arXiv Preprint arXiv:1701.04862*.

Arjovsky, M., Chintala, S., & Bottou, L. (2017). Wasserstein generative adversarial networks. *Proceedings of the 34th International Conference on Machine Learning*, 214–223.

Basak, S., Kar, S., Saha, S., Khaidem, L., & Dey, S. R. (2019). Predicting the direction of stock market prices using tree-based classifiers. *The North American Journal of Economics and Finance*, *47*, 552–567.

Bhuriya, D., Kaushal, G., Sharma, A., & Singh, U. (2017). Stock market predication using a linear regression. *International Conference of Electronics, Communication and Aerospace Technology (ICECA)*, *2*, 510–513.

Box, G. E., Jenkins, G. M., Reinsel, G. C., & Ljung, G. M. (2015). *Time series analysis: Forecasting and control*. John Wiley & Sons.

Bustos, O., & Pomares-Quimbaya, A. (2020). Stock market movement forecast: A systematic review. *Expert Systems with Applications*, *156*, 113464. https://doi.org/10.1016/j.eswa.2020.113464

Cao, H., Lin, T., Li, Y., & Zhang, H. (2019). Stock Price Pattern Prediction Based on Complex Network and Machine Learning. *Complexity*, *2019*, 1–12.

Cao, J., & Wang, J. (2019). Stock price forecasting model based on modified convolution neural network and financial time series analysis. *International Journal of Communication Systems*, *32*(12), 1–13.

Chen, S., & Ge, L. (2019). Exploring the attention mechanism in LSTM-based Hong Kong stock price movement prediction. *Quantitative Finance*, *19*(9), 1507–1515. https://doi.org/10.1080/14697688.2019.1622287

Cho, K., Van Merriënboer, B., Gulcehre, C., Bahdanau, D., Bougares, F., Schwenk, H., & Bengio, Y. (2014). Learning phrase representations using RNN encoder-decoder for statistical machine translation. *arXiv Preprint arXiv:1406.1078*.

Chong, E., Han, C., & Park, F. C. (2017). Deep learning networks for stock market analysis and prediction: Methodology, data representations, and case studies. *Expert Systems with Applications*, *83*, 187–205.

Di Persio, L., & Honchar, O. (2017). Recurrent neural networks approach to the financial forecast of Google assets. *International Journal of Mathematics and Computers in Simulation*, *11*, 7–13.

Ebadati, O. M. E., & Mortazavi, M. T. (2018). An efficient hybrid machine learning method for time series stock market forecasting. *Neural Network World*, *28*(1), 41–55.

Fenghua, W. E. N., Jihong, X., Zhifang, H. E., & Xu, G. (2014). Stock price prediction based on SSA and SVM. *Procedia Computer Science*, *31*, 625–631.

Fischer, T., & Krauss, C. (2018). Deep learning with long short-term memory networks for financial market predictions. *European Journal of Operational Research*, *270*(2), 654–669.

Goodfellow, I. J., Pouget-Abadie, J., Mirza, M., Xu, B., Warde-Farley, D., Ozair, S., Courville, A., & Bengio, Y. (2014). Generative adversarial nets. *Advances in Neural Information Processing Systems*, 2672–2680.

Gu, Q., & Dai, Q. (2021). A novel active multi-source transfer learning algorithm for time series forecasting. *Applied Intelligence*, *51*(3), 1326–1350. https://doi.org/10.1007/s10489-020-01871-5

Gulrajani, I., Ahmed, F., Arjovsky, M., Dumoulin, V., & Courville, A. C. (2017). Improved training of Wasserstein GANs. *Advances in Neural Information Processing Systems*, *30*, 5769–5779.

Gunduz, H. (2021). An efficient stock market prediction model using hybrid feature reduction method based on variational auto encoders and recursive feature elimination. *Financial Innovation*, *7*(1), 1–24. https://doi.org/10.1186/s40854-021-00243-3

Hochreiter, S., & Schmidhuber, J. (1997). Long short-term memory. *Neural Computation*, *9*(8), 1735–1780.

Jin, Z., Yang, Y., & Liu, Y. (2020). Stock closing price prediction based on sentiment analysis and LSTM. *Neural Computing and Applications*, *32*(13), 9713–9729. https://doi.org/10.1007/s00521-019-04504-2

Kodali, N., Abernethy, J., Hays, J., & Kira, Z. (2017). On convergence and stability of GANs. *arXiv Preprint arXiv:1705.07215*.

Kumbure, M. M., Lohrmann, C., Luukka, P., & Porras, J. (2022). Machine learning techniques and data for stock market forecasting: A literature review. *Expert Systems with Applications*, *197*, 116659.

Lin, H. C., Chen, C., Huang, G. F., & Jafari, A. (2021). Stock price prediction using generative adversarial networks. *Journal of Computer Science*, *17*, 188–196. https://doi.org/10.3844/jcssp.2021.188.196

Mirza, M., & Osindero, S. (2014). Conditional generative adversarial nets. *arXiv Preprint arXiv:1411.1784*.

Nasiri, H., & Ebadzadeh, M. M. (2022). MFRFNN: Multi-Functional Recurrent Fuzzy Neural Network for Chaotic Time Series Prediction. *Neurocomputing*, *507*, 292–310.

Nguyen, T.-T., & Yoon, S. (2019). A novel approach to short-term stock price movement prediction using transfer learning. *Applied Sciences*, *9*(22), 4745. https://doi.org/10.3390/app9224745

Pai, P. F., & Lin, C. S. (2005). A hybrid ARIMA and support vector machines model in stock price forecasting. *Omega*, *33*(6), 497–505.

Patel, J., Shah, S., Thakkar, P., & Kotecha, K. (2015). Predicting stock and stock price index movement using trend deterministic data preparation and machine learning techniques. *Expert Systems with Applications*, *42*(1), 259–268.

Salimans, T., Goodfellow, I., Zaremba, W., Cheung, V., Radford, A., & Chen, X. (2016). Improved techniques for training GANs. *Proceedings of the 30th International Conference on Neural Information Processing Systems*, 2234–2242.

Sedighi, M., Jahangirnia, H., Gharakhani, M., & Farahani Fard, S. (2019). A novel hybrid model for stock price forecasting based on metaheuristics and support vector machine. *Data*, *4*(2), 75.

Sinaga, F. M., Jonas, M., & Halim, A. (2019). Stock trend prediction using SV-kNNC and SOM. *Fourth International Conference on Informatics and Computing (ICIC)*, 1–5. https://doi.org/10.1109/ICIC47613.2019.8985731

Singh, R., & Srivastava, S. (2017). Stock prediction using deep learning. *Multimedia Tools and Applications*, *76*(18), 18569–18584.

Sonkiya, P., Bajpai, V., & Bansal, A. (2021). Stock price prediction using BERT and GAN. *arXiv Preprint arXiv:2107.09055*.

Villani, C. (2009). *Optimal transport: Old and new*. Springer.

Wang, H.-Y., Li, H., & Shen, J.-Y. (2019). A novel hybrid fractal interpolation-SVM model for forecasting stock price indexes. *Fractals*, *27*(04), 1950055. https://doi.org/10.1142/S0218348X19500555

Xu, Y., Chhim, L., Zheng, B., & Nojima, Y. (2020). Stacked deep learning structure with bidirectional long-short term memory for stock market prediction. *International Conference on Neural Computing for Advanced Applications*, 447–460. https://doi.org/10.1007/978-981-15-7670-6_37

Zhang, K., Zhong, G., Dong, J., Wang, S., & Wang, Y. (2019). Stock market prediction based on generative adversarial network. *Procedia Computer Science*, *147*, 400–406. https://doi.org/10.1016/j.procs.2019.01.256

Zhang, W., Zhang, S., Zhang, S., Yu, D., & Huang, N. (2019). A novel method based on FTS with both GA-FCM and multifactor BPNN for stock forecasting. *Soft Computing*, *23*(16), 6979–6994.

Zhou, X., Pan, Z., Hu, G., Tang, S., & Zhao, C. (2018). Stock Market Prediction on High-Frequency Data Using Generative Adversarial Nets. *Mathematical Problems in Engineering*, *2018*, 1–11.